\documentclass{jetpl}
\twocolumn

\lat

\title{Experimental Investigation of the Dark Pseudoresonance
on the $D_1$ Line of the $^{87}Rb$ Atom Excited by a Linearly
Polarized Field}

\rtitle{Experimental Investigation of the Dark Pseudoresonance
\ldots}

\sodtitle{Experimental Investigation of the Dark Pseudoresonance
on the $D_1$ Line of the $^{87}Rb$ Atom Excited by a Linearly
Polarized Field}

\author{S. A. Zibrov$^{a,b}$, V. L. Velichansky$^{a,b}$, A. S. Zibrov$^{a,e}$, A. V.
Ta$\breve{i}$chenachev$^{c,d}$, and V. I. Yudin$^{c,d}$}

\rauthor{S. A. Zibrov, V. L. Velichansky, A. S. Zibrov, A. V.
Ta$\breve{i}$chenachev, and V. I. Yudin}

\sodauthor{Zibrov, Velichansky, Zibrov, Ta$\breve{i}$chenachev,
and Yudin}

\address{$^{a}$ Lebedev Physical Institute, Russian Academy of Sciences, Leninsky
pr. 53, Moscow, 117924, Russia,\\
$^{b}$ Moscow Engineering Physics Institute (State University),
Kashirskoe sh. 31, Moscow, 115409 Russia\\
e-mail: szibrov@yandex.ru\\
$ ^{c}$ Institute of Laser Physics, Siberian Division, Russian
Academy of Sciences, pr. Lavrent'eva 13/3, Novosibirsk, 630090
Russia\\
$^d$ Novosibirsk State University, ul. Pirogova 2, Novosibirsk,
630090 Russia\\
 $^e$ Department of Physics,
Harvard University, Cambridge and Harvard – Smithsonian Center for
Astrophysics, Cambridge, Massachusetts 02138, USA}

\dates{22 August 2005}

\abstract{The measurements of the metrological characteristics
(amplitude, width, and shift in the magnetic field) of the dark
pseudoresonance, which was proposed by Kazakov et al.
[quant-ph/0506167] as the reference resonance for an atomic
frequency standard, are reported. It has been shown that the
characteristics of the pseudoresonance are worse than those of the
unsplit electromagnetically induced transparency resonance for the
excitation scheme with the $lin||lin$ polarization on the $D_1$
line of the $^{87}$Rb atom. }

\PACS{42.50.Gy, 42.65.–k, 42.65.Ky}

\begin{document}

\maketitle

Since the 1970s, two-photon resonances free of Doppler broadening
have been successfully used as a reference for quantum frequency
standards \cite{1,2}. In 1993, the effect of coherent population
trapping was proposed to be used to create a microwave frequency
standard based exclusively on optical elements without a microwave
cavity \cite{3}. In recent years, the possibility of creating an
atomic clock based on this effect is actively analyzed
\cite{4,5,6,7}. Two copropagating laser fields acting on the
allowed electric dipole transitions in the $\Lambda$-
configuration create a long-lived superposition of states in
hyperfine sublevels of the ground states of the alkali-metal
atoms. When the difference between the frequencies varies near the
hyperfine splitting frequency $\Delta_{hfs}$ , the transmission
resonance is observed (named as coherent population trapping
resonance or $\Lambda$- resonance). The resonance width in the
limit of low intensities is determined by the coherence lifetime
in the ground state. To date, a frequency stability of 6.4$\times
$$10^{-13}$ ( $\tau $ = 2000 s) is reached in such a microwave
standard \cite{6}. Moreover, the possibility of a radical decrease
in the volume of quantum discriminators of the coherent population
trapping clock on Cs and Rb (to about 10 mm$^3$ ) was demonstrated
\cite{8,9}. Investigations in this direction are actively
continued.

It is known that the stability of quantum frequency standards
increases with increasing the amplitude $A $  of the resonance and
with decreasing its width $W$ (\cite{5}, Eq. (8)):

\begin{equation}\label{eq1}
    \sigma(\tau)=
{\sqrt{\eta I_{bg}}}
{\frac{1}{\Delta_{hfs}}{\frac{W}{A}}{\tau^{-1/2}}}
\end{equation}

\begin{figure}
\centering
 \includegraphics[width=7cm]{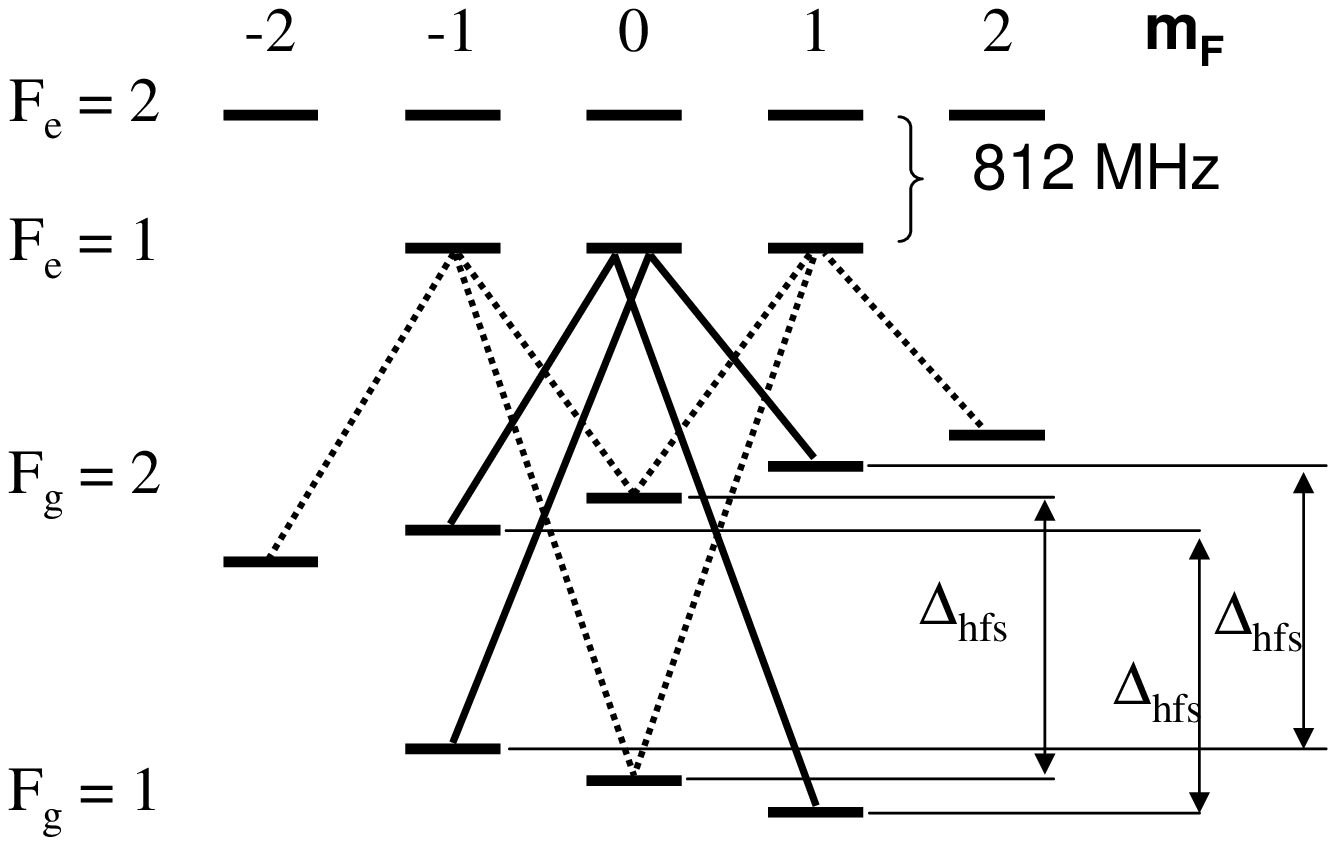}
   \includegraphics[width=8cm]{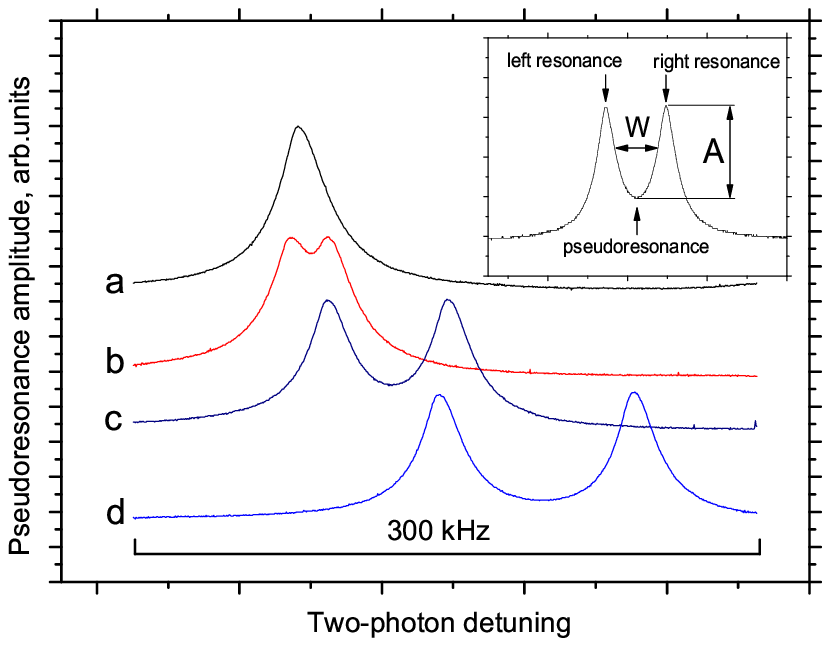}
\caption{ \label{1} Fig.1:~(Upper panel) The scheme of the $lin ||
lin$ excitation of the two-photon transition ($F = 1\rightarrow F
= 2$) $5S_{1/2}$ of the $D_1$ line of the $^{87}$Rb atom. The
lower panel illustrates the appearance of the pseudoresonance as
the longitudinal magnetic field increases: (a) 0.35, (b) 3.0, (c)
9.0, and (d) 15 G; A and $W$ are the amplitude and width of the
resonance, respectively. The total power of the laser radiation
was equal to 2 mW and the beam diameter was equal to 4 mm.
 }
\end{figure}

where $\sigma(\tau)$ is the Allan parameter, $\Delta_{hfs}$ is the
standard frequency, and $I_{bg}$ is the background caused by
radiation that is not absorbed by the medium. For this reason, the
search for schemes for the excitation of the coherent population
trapping resonance with high contrast ($>$5\%), minimum width, and
light shift is of current interest for an increase in the
stability of the clock.

A "push–pull" pumping scheme that allows the production of a pure
coherent state was proposed in \cite{9}. Atoms in such a state do
not interact with the field; i.e., the atom-field interaction
operator is equal to zero  $ -\hat{d}{E}|{Dark}\rangle=0$, where
$|{Dark}\rangle$ is the coherent superposition of the wave
functions of the Zeeman sublevels of the ground state. In that
work, a contrast of about 30\% was experimentally achieved. The
pure dark state prepared by means of a standing wave was also
demonstrated in \cite{10}. In \cite{11}, it is shown that, when
the red detuning of the frequency of the pump field is equal to
the hyperfine splitting $\Delta_{hfs}$ of the ground state, a
contrast of about $20\%$ is reached with almost zero shifts. We
note that, although this work does not involve the coherent
population trapping effect, it is close to the subject under
discussion in its aim: the use of the bichromatic field, and the
$\Lambda$-configuration of the involved processes. In order to
improve the metrological characteristics of the coherent
population trapping resonance, the pulse scheme of detecting the
resonance was studied by the Ramsey method \cite{12}. A unique
possibility was pointed out in \cite{13} for forming the pure
coherent population trapping resonance (free of trap states) due
to the interaction between the bichromatic field with $lin || lin$
polarized components and the 5$P_{1/2}$ state of $^{87}$Rb atoms.
In that case, a pure dark state appeared under the action of the
linearly polarized bichromatic field under the condition of the
spectral resolution of the hyperfine structure of the excited
state (Fig.\ref{1}). The achievement of a contrast of about 50\%
was reported in that work. It was pointed out that the magnetic
field sensitivity (quadratic Zeeman shift) of the proposed
resonance should be 1/1.33 of the shift of the resonance formed
due to excitation by the circularly polarized bichromatic field
($\sigma^{+}$-$\sigma^{+}$ scheme). In that work, it was also
noted that the resonance was split into two resonances in high
magnetic fields. This splitting arises because the g factors of
two hyperfine sublevels of the ground state are slightly different
due to the nuclear-spin contribution.


\begin{figure}
\centering
\includegraphics[height=5cm]{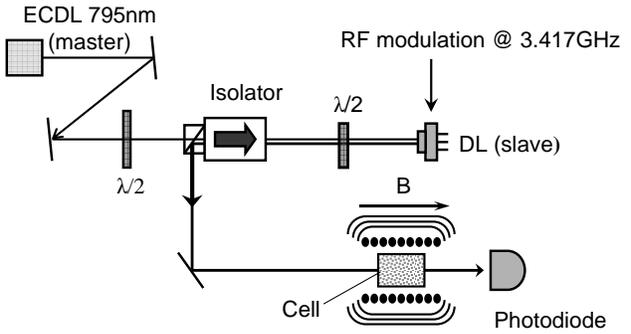}
\caption{ \label{2} Fig.2:~Layout of the setup: the ECDL is the
external-cavity injection laser, the DL is the diode laser, the
Isolator suppresses  reflections on 60 dB. The cell temperature
was equal to $50^0$ C.}
\end{figure}

In \cite{14}, it was proposed to use a dip arising due to  a
splitting of the resonance as the reference for the microwave
standard (see Fig.1). The authors of \cite{14} referred to the dip
as pseudoresonance, because it appeared due to the splitting of
the coherent population trapping resonance in the magnetic field.
The estimates made in that work showed the possibility of reaching
a frequency stability of $10^{-14}$/$\sqrt{\tau}$ . Such a high
stability would enable one to consider the pseudoresonance as a
high-priority and promising tool for creating atomic frequency
standards based on the coherent population trapping effect. For
this reason, theoretical and experimental investigation of the
pseudoresonance, as well as comparison with the unsplit coherent
population trapping resonance, seems to be of interest.

In this paper, the experimental results on certain metrological
characteristics of the pseudoresonance are reported. It is shown
that they are worse than those of the initial (unsplit) coherent
population trapping resonance.

Figure 2 shows the layout of the experimental setup that consists
of a laser system, a cell filled with $^{87}$Rb vapor, and a
detection system. The experiment was conducted with a Pyrex
cylindrical cell (40 mm in length and 25 mm in diameter)
containing Ne at a pressure of 4 Torr and isotopically pure
$^{87}$Rb. The cell was placed inside a solenoid, which provided
variation in the longitudinal magnetic field. To screen the
external laboratory field, the cell was placed inside three
cylindrical magnetic screens. The heating of the cell was
performed by means of a bifilar nichrome wire coiled around the
inner magnetic screen. The cell temperature was equal to
$50^{0}$C. The bichromatic resonance field was produced by
modulating the current of a ``slave'' laser  whose frequency was
 matched by the frequency of a single-mode
external-cavity injection laser (ECLD, ``master''). To this end,
the radiation of the latter laser was injected through a isolator
into the active region of the driven laser (DL). In this case, the
modulation did not disturb  the regime of the maser laser (ECLD).
The injection current of the driven laser (DL) was modulated at a
frequency of $\Delta_{hfs}$/2 = 3.417 GHz by means of an Agilent
E8257D-502 microwave generator, which was connected to the
``slave'' laser through a Minicircuits ZFBT-6G T bias. Such a
procedure ensured the generation of resonant optical fields with a
high correlation degree of phase noises. The ratio of the
intensities of these fields could be changed by slightly varying
the current of the driven laser. The resonant fields carried
approximately 50\% of the total power of the radiation (2 mW). The
ratio of the intensities of the resonant fields was equal to 1.4,
and the amplitude of the coherent population trapping resonance
was maximal. The remaining power was contained in the carrier and
higher order side frequencies. The laser beam in the cell had a
diameter of 4 mm. The coherent population trapping resonance was
excited by the linearly polarized first-order components, which
were tuned to the $F_g = 1 \leftrightarrow F_e = 1$ and $F_g = 2
\leftrightarrow F_e = 1$ transitions. The intensity of the
radiation passed through the cell was measured by a photodiode. In
order to study the pseudoresonance, the modulation frequency
$\Delta_{hfs}/2$ was linearly scanned in a narrow range ($\sim150
kHz$) for various magnetic fields. The amplitude and width (Fig.3)
of the resonance, as well as its amplitude-to-width ratio and its
position (Fig.4), were studied as functions of the magnetic field.

\begin{figure}
\centering
 \center\includegraphics[width=8.0cm]{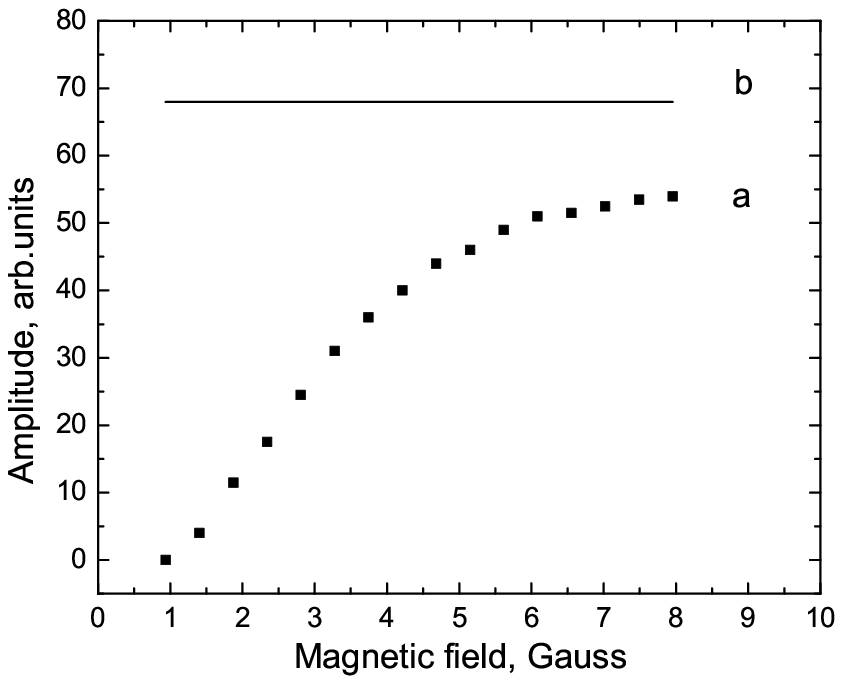} \includegraphics[width=8.0cm]{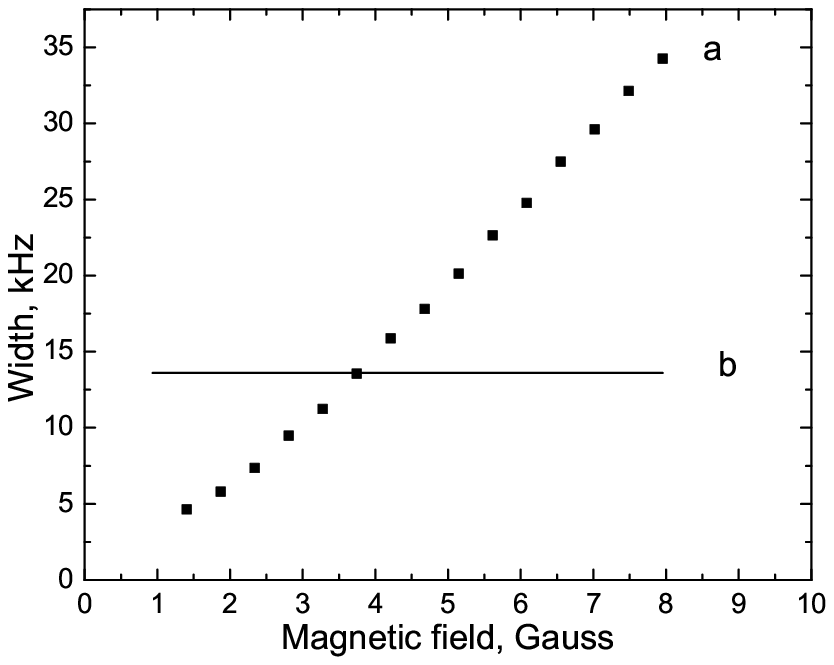}
\caption{\label{3} Fig.3:~ Magnetic-field dependence of the
amplitude $A$ (upper panel) and  width $W$ (lower panel) of
pseudoresonance, where the horizontal straight lines show the
respective values of the unsplit coherent population trapping
resonance in the presence of a magnetic field of 0.2 G. }
\end{figure}

In the absence of the magnetic field, the amplitude of the
coherent population trapping resonance is one order of magnitude
less than the amplitudes of the resonances observed in the
presence of the magnetic field. In the presence of the magnetic
field, this resonance is split into three resonances: the
magnetically independent central resonance at the transition
frequency $\Delta_{hfs}$ and two magnetically dependent
resonances. The central resonance is formed by two $\Lambda$
transitions: \{$| F_g = 1, m_F = -1\rangle \leftrightarrow | F_e =
1, m_F = 0\rangle \leftrightarrow |F_g = 2, m_F = 1\rangle $\} and
\{$|F_g = 2, m_F = -1\rangle \leftrightarrow |F_e = 1, m_F = 0
\rangle \leftrightarrow |F_g = 1, m_F = 1 \rangle$\}. The main
contribution to the magnetically dependent resonances comes from
the following $\Lambda$ transitions: \{$| F_g = 2, m_F = -1\rangle
\leftrightarrow |F_e = 1, m_F = 0 \rangle \leftrightarrow | F_g =
1, m_F = -1 \rangle$\} and \{$| F_g = 2, m_F = +1
\rangle\leftrightarrow | F_e = 1, m_F = 0 \rangle \leftrightarrow
|F_g = 1, m_F = 1\rangle $\}. Immediately after the appearance of
the magnetic field, the amplitude of the central resonance
increases and its contrast reaches 40\%. The amplitude of the
magnetically dependent resonances also increases to 12\%. The
resonances grow upon the appearance of the magnetic field, because
the removal of degeneration destroys dark trap states on the
Zeeman sublevels that belong to the same hyperfine level and on
which atoms "are hidden." To destroy these traps, a magnetic field
whose magnitude is higher than the width of the resonance is
necessary. In not too strong fields, the width of the resonance is
determined by the optical pumping rate for the ground state
\cite{15}. A further increase in the magnetic field (to 0.5 G, see
Fig.5) results in a new splitting of the central resonance. In
this way, the pseudoresonance appears. The amplitude of the
pseudoresonance is saturated at a magnetic field exceeding 7.0 G
but does not reach the amplitude of the initial coherent
population trapping resonance, see Fig.3(upper panel). The width
of the pseudoresonance becomes less than the width of the initial
resonance at a magnetic field of less than 4.0 G, for which the
amplitude of the pseudoresonance is very small.

It is seen in Fig.4 that the maximum A/W ratio is reached at a
magnetic field of about 3.4G. For this field both the width and
amplitude of the pseudoresonance are less than the respective
values of the initial resonance. The ratio A/W for the
pseudoresonance is worse at any magnetic field. It is worth noting
that, for locking of the quartz-oscillator frequency in the atomic
clock, resonances with a sharp peak (with a larger slope of the
first derivative) are preferable over those with a smooth peak, as
in the case of the pseudoresonance.

\begin{figure}
\centering \center\includegraphics[width=8cm]{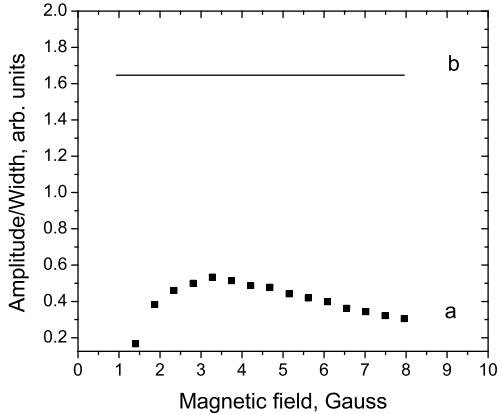}
\caption{\label{4} Fig.4:~ Magnetic-field dependence of the
amplitude-to-width ratio for the pseudoresonance, where the
horizontal straight line shows the ratio for the unsplit coherent
population trapping resonance in the presence of a magnetic field
of 0.2 G. }
\end{figure}

We point to one more feature of the behavior of the split
resonance. Figure 5 shows the shift of  two true coherent
population trapping resonances (lines a and c) and  the
pseudoresonance (line b) as the longitudinal field varies. This
dependence for the coherent population trapping resonances has the
form (see, e.g., \cite{14})

\begin{figure}
\centering \center\includegraphics[width=8cm]{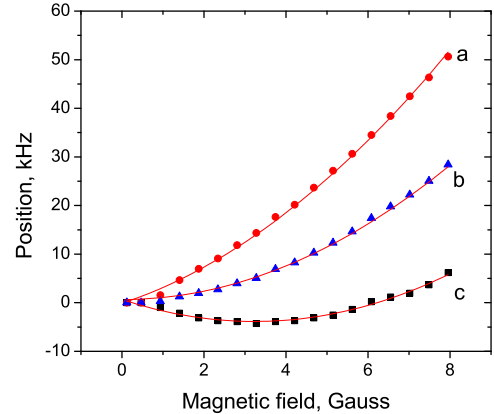}
\caption{\label{5}Fig.5:~Magnetic-field dependence of the position
of the (a) right resonance, (b) pseudoresonance, and (c) left
resonance. The origin of the abscissa axis corresponds to the
position of the unsplit coherent population trapping resonance.
The points are experimental data and lines are approximations. }
\end{figure}

\begin{equation}\label{eq2}
    \Delta=\Delta_{hfs} + \frac{2\textrm{g}_I \mu_N }{\hbar}H +
    \frac{3\textrm{g}^2_J
    \mu_B^2}{8\omega_{hfs}\hbar^2}H^2
\end{equation}

where $\mu_B$ is the Bohr magneton, $\mu_ N$ is the nuclear
magneton, and $\textrm{g}_I$ and $\textrm{g}_J$ are the nuclear
and electron Lande factors, respectively. It is seen that the
shift of the coherent population trapping resonances are the sum
of the linear and quadratic contributions from both transitions
$m_F = -1 \leftrightarrow m_F = +1$ (left resonance) and $m_F = +1
\leftrightarrow m_F = -1$ (right resonance) involved in the
formation of the reference resonance. For low magnetic fields, the
left resonance is shifted from the right resonance (and from the
pseudoresonance) according to the linear law. However, as the
field increases, the left resonance changes the direction of its
shift and begins to move in the same direction as the right
resonance, because the quadratic term begins to dominate in Eq.
(2). From experimental line b , the quadratic dependence of the
position of the dark resonance on the magnetic field is found with
a coefficient of about $0.43\pm 0.04kHz/G^2$. As was predicted in
[13], this coefficient is less than the corresponding coefficient
for the standard atomic clock by a factor of 1.33 \cite{16}.

In this work, certain metrological characteristics of the dark
pseudoresonance have been experimentally studied. The results
provide the conclusion that these characteristics are noticeably
worse than the respective characteristics of the initial (unsplit)
coherent population trapping resonance from which the
pseudoresonance appears. Thus, we think that the use of the
pseudoresonance as the reference for the atomic clock is not an
optimum solution when using the $lin || lin $ excitation scheme on
the $D_1$ line of the $^{87}Rb$ atom. At least more detailed
theoretical investigation is required for determining the
experimental conditions (cell sizes, buffer gas pressure, etc.)
under which the pseudoresonance could be preferable over the
unsplit coherent population trapping resonance

We are grateful to Tamara Zibrova for the high-quality blooming of
the mirrors of the laser diodes. The work of A.V.T. and V.I.Yu.
was supported by INTAS (grant no. 01-0855) and the Russian
Foundation for Basic Research (project nos. 05-02-17086 and 04-02-
16488).

\end{document}